\documentclass[12pt,aps,superscriptaddress,nofootinbib,preprintnumbers]{revtex4}
\usepackage{graphicx}
\usepackage{slashed}

\newcommand{\ba}{\begin{array}}
\newcommand{\ea}{\end{array}}
\newcommand{\bd}{\begin{displaymath}}
\newcommand{\ed}{\end{displaymath}}
\newcommand{\be}{\begin{equation}}
\newcommand{\ee}{\end{equation}}
\def\bt{\begin{table}}
\def\et{\end{table}}
\def\bc{\begin{center}}
\def\ec{\end{center}}
\def\bi{\begin{itemize}}
\def\ei{\end{itemize}}
\def\bw{\begin{widetext}}
\def\ew{\end{widetext}}

\def\bea{\begin{eqnarray}}
\def\eea{\end{eqnarray}}
\def\beas{\begin{eqnarray*}}
\def\eeas{\end{eqnarray*}}

\def\N0{\widetilde{\chi}^0}

\begin{document}

\setlength{\parskip}{0.1cm}

\preprint{OSU-HEP-13-04}
\preprint{RECAPP-HRI-2013-014}

\vspace*{0.8in}

\title{New Signals for Doubly--Charged Scalars and Fermions\\[-0.1in]
at the Large Hadron Collider }

\author{K.S. Babu}
\email{kaladi.babu@okstate.edu}

\author{Ayon Patra}
\email{ayon@okstate.edu}
\affiliation{Department of Physics, Oklahoma State University, Stillwater, OK
74078, USA}

\author{Santosh Kumar Rai}
\email{skrai@hri.res.in}
\affiliation{Regional Centre for Accelerator-based Particle Physics,\\[-0.08in]
Harish-Chandra Research Institute, Chhatnag Road, Jhusi,
Allahabad 211019, India}


\begin{abstract}

\begin{center}
{\large{Abstract}}
\end{center}

Several extensions of the Standard Model have light doubly-charged Higgs bosons in their particle spectrum. The supersymmetric versions of these models introduce fermionic superpartners of these doubly-charged Higgs bosons, the Higgsinos, which also remain light. In this work we analyze a new collider signal resulting from the pair production and decay of a light doubly-charged Higgsino to an even lighter doubly-charged Higgs boson.
We focus on the minimal left-right supersymmetric model with automatic $R$-parity conservation, which predicts such a light doubly-charged Higgs boson and its Higgsino partner at the TeV scale, which are singlets of $SU(2)_L$.  We investigate the distinctive signatures of these particles with four leptons and missing transverse energy in the final state at the Large Hadron Collider and show that the discovery reach for both particles can be increased in this channel.
\end{abstract}
\maketitle

\newpage
\section {Introduction}

Several extensions of the Standard Model (SM) predict the existence of doubly-charged Higgs bosons.  In some cases these particles remain light,
which motivates searches for them in high energy collider experiments.  The minimal left-right supersymmetric model with automatic $R$-parity
conservation is an example, where a light doubly-charged Higgs boson arises as a pseudo-Goldstone boson of the $SU(2)_R$ gauge symmetry breaking
\cite{lrsusy1,aulakh,Chacko:1997cm,lrsusy2}.
Models with radiative neutrino mass generation \cite{radneu}, Type-II see-saw mechanism \cite{type2seesaw} for small neutrino masses,
and the  3-3-1 model \cite{331 models} are some other examples of SM extensions which have doubly-charged Higgs bosons. Supersymmetric versions of these models also have doubly-charged Higgsinos, which are the fermionic partners of the Higgs bosons. If the doubly-charged Higgs boson is light,  its Higgsino partner cannot be much heavier and must have mass of the order a few hundred GeV to a few TeV, in the
context of low energy supersymmetry (SUSY).

In this paper we study a new signal for the doubly-charged Higgs bosons and Higgsinos in SUSY models which arises through the pair-production of the doubly-charged Higgsinos.  Each  Higgsino decays into a doubly-charged Higgs boson and the lightest supersymmetric particle (LSP) which escapes
detection.  Thus the final state would have four leptons and missing transverse energy, with the same-sign dileptons originating
from the decays of the doubly-charged Higgs bosons showing characteristic peaks in the invariant mass distribution.  We show by detailed
calculations in the context of left-right supersymmetric model that the reach at the LHC for both these doubly-charged particles can be enhanced by
studying this mode.  While we focus on the minimal supersymmetric left-right model, these new signals should also be present in other SUSY models with a light doubly-charged Higgsino and a lighter doubly-charged Higgs boson.

The focus of our analysis will be the minimal supersymmetric left-right gauge model.
Left-right symmetric models \cite{lrs} have a number of attractive features which are not naturally present in the Standard Model. Firstly, it explains the small neutrino masses through the see-saw mechanism \cite{seesaw} in a compelling manner  -- unlike the SM, existence of right-handed neutrinos is required by gauge symmetry here.  Secondly, it provides a natural understanding of the origin of parity violation as a spontaneous phenomenon \cite{lrs}.  Thirdly, with the inclusion of supersymmetry, this model solves the gauge hierarchy problem and in its simplest version, also provides an automatic $R$-parity.  This symmetry arises as remnant of the $(B-L)$ gauge symmetry \cite{rnm} and leads to a stable light supersymmetric particle which can be a candidate for dark matter. With supersymmetry these models also provide natural solutions to the strong CP problem and the SUSY CP problem \cite{bdm}.

In the minimal left-right supersymmetric model, the gauge group is extended to $ G_{3221} = {SU(3)}_c \times {SU(2)}_L \times {SU(2)}_R \times {U(1)}_{B-L} $. The ${SU(2)}_R \times {U(1)}_{B-L} $ symmetry breaks at a high scale resulting in most of the new particles getting very heavy masses. The right-handed neutrino mass is at this scale and facilitates the generation of the light neutrino mass via the see-saw mechanism. The doubly-charged Higgs supermultiplet, on the other hand, remains light and can produce new signals which is the focus of our analyze in this paper.

To understand why the doubly-charged Higgs boson remains light in the minimal model, we need to look at the symmetry breaking sector. To spontaneously break the $SU(2)_R$ gauge symmetry and to generate large Majorana mass for the right-handed neutrino, we need to introduce a Higgs multiplet with quantum numbers $(1,1,3,-2)$ under the group $G_{3221}$. This right-handed triplet contains three complex fields:  a doubly-charged, a singly-charged and a neutral field denoted by $\delta^{c^{--}}, \delta^{c^-}, \delta^{c^0}$ respectively. The $\delta^{c^-}$ and the phase of $\delta^{c^0}$ are absorbed by the gauge fields via the super-Higgs mechanism to generate masses for the $W_R^{\pm}$ and $Z_R$ gauge bosons. The real part of $\delta^{c^0}$ gets a mass through the Higgs potential.  The $\delta^{c^{--}}$ field, on the other hand, is not absorbed by any gauge bosons, nor does it acquire a mass from the superpotential of the minimal model.  Thus it behaves like pseudo-Goldstone boson, acquiring its mass only after supersymmetry breaking.\footnote{ The superpotential of the model, which only has quadratic mass terms, has an enhanced global $U(3,c)$ (complexified $U(3)$) symmetry which is broken to an $U(2,c)$ by the VEV of this Higgs multiplet. This leads to five massless superfields of which three are absorbed to give mass to the heavy gauge bosons and the remaining are the two doubly-charged Higgs bosons. Since SUSY is unbroken at this stage, the doubly-charged Higgsino is degenerate with
the doubly-charged Higgs boson.} As a result, the right-handed doubly-charged Higgs bosons and the doubly-charged Higgsinos remain light in this model.

The doubly-charged Higgs bosons decay to two same charge leptons, which can be seen relatively easily in collider experiments via the invariant mass peak in the dilepton mass spectrum. LHC has been looking for signals of doubly-charged Higgs boson in the four lepton final states \cite{LHCdcbexp1,LHCdcbexp2}. The experimental lower limit inferred on the mass of such Higgs bosons would depend on the assumed branching
ratios into leptons of definite flavors.  For example, CMS experiment quotes a 95\% CL lower limit of 355 GeV for the mass of a doubly-charged
Higgs boson arising from an $SU(2)_L$ triplet, if it decays with equal branching ratios of 33\% into $e^+ e^+$, $\mu^+ \mu^+$ and
$\tau^+ \tau^+$.  The 95\% CL lower limit on such a Higgs particle from the ATLAS experiment is 318 GeV.  These limits are somewhat weaker
for an $SU(2)_L$ singlet doubly-charged Higgs boson, since its production cross section is smaller compared to the case when
it is a $SU(2)_L$ triplet.  For example, ATLAS collaboration quotes a lower limit on the mass of an $SU(2)_L$ singlet doubly-charged scalar
that decays with a 33\% BR into $\mu^+ \mu^+$ of about 220 GeV, while the limit is about 210 GeV if it decays into $e^+ e^+$ with the
same branching ratio.  We anticipate that the lower limit, when both modes are combined, would be somewhat smaller than 300 GeV,
for an $SU(2)_L$ singlet, as in our case.\footnote{When an $SU(2)_L$ singlet doubly-charged Higgs boson decays 100\% of the time into
$\mu^+ \mu^+$ (or $e^+e^+$), the ATLAS lower limit on its mass is about 310 (or 320) GeV \cite{LHCdcbexp2}.}

The decay of doubly-charged Higgsino $(\widetilde{\delta}^{c^{\pm \pm}})$ through a doubly-charged Higgs boson $(\delta^{c^{\pm \pm}})$ can produce new signals through the following process:
\begin{center}
 $\widetilde{\delta}^{c^{\pm \pm}} \rightarrow \delta^{c^{\pm \pm}} \tilde{\chi}_1^0 \rightarrow l^\pm l^\pm \tilde{\chi}_1^0 ~.$
\end{center}
So the pair production of doubly-charged Higgsinos yields a final state consisting of four leptons and missing transverse energy due to the LSP escaping the detector. This process, which has not been explored before to the best of our knowledge, gives a unique collider signature which can help improve the discovery reach of doubly-charged particles. The invariant mass plot would show a peak at the doubly-charged Higgs mass for the same-sign lepton while there would be no such peak for opposite-sign leptons.
The angular distributions for the final state leptons also
show a peak at a low value of $\Delta R$ (defined later in the paper) for same-sign leptons while the opposite-sign leptons have a peak at a much higher value.
Using these distributions we can probe deeper into the model than one could just by looking at the pair production of the doubly-charged Higgs bosons. The cross section for pair production of doubly-charged Higgsinos is larger compared to the cross section for
the pair production of doubly-charged Higgs bosons of the same mass.
From the current data at the LHC, we expect around 30 events for the process discussed in this paper, if the doubly-charged Higgs boson
has a mass of about 500 GeV, and if it decays into a doubly-charged Higgs boson of mass around 300 GeV.

In section~\ref{sec:LRSUSY} we describe the model and the Lagrangian needed for our analysis. We also explain the origin of masses of the doubly-charged Higgs boson and the Higgsino and show that they remain light. In section~\ref{sec:signals}, we present our analysis of the production and decay of the doubly-charged scalars and fermions and give the collider signatures which can be observed at the LHC. Section~\ref{sec:conclusion} gives a discussion of the results that we have obtained and how we can distinguish our signal against the background.

\section {A brief review of the Left-Right Supersymmetric Model}
\label{sec:LRSUSY}

In this section, we briefly review the relevant features of the minimal supersymmetric left-right model (LRSUSY) necessary for the analysis which follows in the later sections \cite{lrsusy1,lrsusy2}.\footnote{For alternative versions of SUSY left-right model, see Ref. \cite{lrsusy3}.}
 The chiral matter in LRSUSY consist of three families of quark and lepton superfields,
\begin{eqnarray}
\!\!Q\!\!&=&\!\!\left (\begin{array}{c}
u\\ d \end{array} \right ) \sim \left (3,2, 1, \frac13 \right ),
Q^c\!=\!\left (\begin{array}{c}
d^c\\-u^c \end{array} \right ) \sim \left ( 3^{\ast},1, 2, -\frac13
\right ),\nonumber \\
L&=&\left (\begin{array}{c}
\nu\\ e\end{array}\right ) \sim\left ( 1,2, 1, -1 \right ),
L^c=\left (\begin{array}{c}
e^c \\ -\nu^c \end{array}\right ) \sim \left ( 1,1, 2, 1 \right ),
\end{eqnarray}
where the numbers in the brackets denote the quantum numbers under
$SU(3)_c \times SU(2)_L \times SU(2)_R \times U(1)_{B-L}$ gauge groups.

The minimal Higgs sector consists of the following superfields:
\begin{eqnarray}
\Delta(1,3,1,2)&=&\left (\begin{array}{cc}
\frac{\delta^{+}}{\sqrt{2}} & \delta^{++}\\ \delta^{0} & -\frac{\delta^{+}}{\sqrt{2}}  \end{array} \right ),~~
\overline{\Delta}(1,3,1,-2)=\left (\begin{array}{cc}
\frac{\overline{\delta}^{-}}{\sqrt{2}} & \overline{\delta}^{0}\\ \overline{\delta}^{--} & -\frac{\overline{\delta}^{-}}{\sqrt{2}}  \end{array} \right ),\nonumber \\
\Delta^{c}(1,1,3,-2)&=&\left (\begin{array}{cc}
\frac{\delta^{c^{-}}}{\sqrt{2}} & \delta^{c^{0}}\\ \delta^{c^{--}} & -\frac{\delta^{c^{-}}}{\sqrt{2}}  \end{array} \right ),~~
\overline{\Delta}^{c}(1,1,3,2)=\left (\begin{array}{cc}
\frac{\overline{\delta}^{c^{+}}}{\sqrt{2}} & \overline{\delta}^{c^{++}}\\ \overline{\delta}^{c^{0}} & -\frac{\overline{\delta}^{c^{+}}}{\sqrt{2}}  \end{array} \right ),\nonumber \\
\Phi_{a}(1,2,2,0)&=&{\left (\begin{array}{cc}
\phi^{+} & \phi^{0}_{2} \\ \phi^{0}_{1} & \phi^{-}_{2} \end{array} \right )}_a  (a=1,2), ~~~S(1,1,1,0)~~.
\end{eqnarray}

The $\Delta^{c}$ and $\overline{\Delta}^{c}$ fields are the right-handed triplets and are necessary for breaking the $SU(2)_R \times U(1)_{B-L}$ symmetry without inducing any $R$-parity violating couplings. The $\Delta$ and $\overline\Delta$ fields are their left-handed partners which are required for parity invariance. The two bidoublets $\Phi_a$ are needed to give mass to the quarks and leptons and to generate the CKM mixings. The singlet $S$ is there to make sure that the $SU(2)_R \times U(1)_{B-L}$ symmetry breaking occurs in the supersymmetric limit \cite{lrsusy2}.

The superpotential of the model is given as
\begin{eqnarray}
W&=& Y_{u} Q^{T} \tau_{2}\Phi_{1}\tau_{2}Q^{c} + Y_{d} Q^{T} \tau_{2}\Phi_{2}\tau_{2}Q^{c} +Y_{\nu} L^{T} \tau_{2}\Phi_{1}\tau_{2}L^{c} +Y_{l} L^{T} \tau_{2}\Phi_{2}\tau_{2}L^{c} \nonumber\\
&+&i(f^{*} L^{T}\tau_{2} \Delta L+f L^{c^{T}}\tau_{2}\Delta^{c}L^{c} ) \nonumber\\
&+&S[Tr(\lambda^{*} \Delta \overline{\Delta}+\lambda \Delta^{c}\overline{\Delta}^{c}) + \lambda_{ab}^{'} Tr(\Phi_{a}^{T}\tau_{2}\Phi_{b}\tau_{2}) -M_{R}^{2}] + W'
\label{sup}
\end{eqnarray}
where
\begin{eqnarray}
W'&=&\left[M_{\Delta} Tr(\Delta\overline{\Delta})+M_{\Delta}^* Tr(\Delta^c\overline{\Delta}^c)\right]+\mu_{ab} Tr\left(\Phi^T_a\tau_2 \Phi_b\tau_2\right)+M_{S}S^2+\lambda_{S}S^3~.
\end{eqnarray}
Here $Y_{u,d}$ and $Y_{\nu,l}$ are the Yukawa couplings for quarks and leptons respectively and $f$ is the Majorana neutrino Yukawa coupling matrix. This is the most general superpotential. $R$-parity is automatically preserved in this case, which is a consequence of $(B-L)$ being
part of the gauge symmetry. Putting $W'=0$ gives an enhanced $U(1)$ \,$R$-symmetry in the theory. Under this $R$-symmetry, $Q,Q^C,L,L^C$ fields have a charge of $+1$, $S$ has charge $+2$ and all other fields have charge zero with $W$ carrying a charge of $+2$. Putting $W'=0$ also helps in understanding the $\mu$-parameter of MSSM since it is induced as $\mu \sim \lambda' \left<S\right>$ from Eq. (\ref{sup}), which is of the scale of SUSY breaking, as necessary.  Setting $W'=0$ would make the doubly-charged left-handed and right-handed Higgsinos degenerate in mass since
both masses are given by $\lambda \left< S\right>$, see Eq. (\ref{sup}).\footnote{Keeping a non-zero $W'$ term does not affect the right-handed particle spectrum, but the left-handed Higgsino becomes very heavy in this case and will not contribute to our new signal. We present results of our analysis with and without the left-handed doubly-charge Higgsino in the light spectrum, so this effect can be disentangled.}

The $SU(2)_R \times U(1)_{B-L}$ symmetry is broken at a large scale by giving a large vacuum expectation value to the right-handed triplet Higgs boson fields $\Delta^{c}$ and $\overline{\Delta}^{c}$. This generates a large right-handed neutrino mass, $M_{\nu^c} = 2 f v_R$,
where $v_R$ is the vacuum expectation value of the $\delta ^{c^0}$ field which breaks the $SU(2)_R$ symmetry. This helps generate a small
Majorana mass for the left-handed neutrino via the see-saw mechanism \cite{seesaw}. The bidoublets get VEVs of the order of electroweak symmetry breaking scale and generate the masses of the quarks and leptons. The singlet $S$ gets a VEV of order the SUSY breaking scale, and helps solve the $\mu$-problem of the MSSM, assuming that  $W' = 0$.

The terms in the Lagrangian which will be most essential for our calculation later are the gauge kinetic terms for the triplet superfields and the quarks and leptons. These terms will give us the interaction vertices between the Higgs boson fields and the gauge bosons as well as the the fermions and the gauge bosons \cite{lagrangian}. The kinetic terms for the triplet scalar fields and the fermions are given by:
\begin{equation}
L = i \sum Tr[\overline{q}_i \slashed{D} q_i]+Tr[(D^\mu \Phi_i)^{\dagger} (D_\mu \Phi_i)]
\end{equation}
where
$q_i = Q,Q^c,\widetilde{\Delta},\widetilde{\overline{\Delta}},\widetilde{\Delta^c},\widetilde{\overline{\Delta}^c}$ and $\Phi_i = {\Delta},{\overline{\Delta}},{\Delta^c},{\overline{\Delta}^c}$.
The covariant derivatives are defined as
\begin{eqnarray}
D_{\mu}Q&=& [\partial_{\mu}-i\frac{g_{L}}{2} \vec{\tau}\cdot\vec{W}_{\mu L}-i\frac{g_V}{6}V_{\mu}]Q \nonumber \\
D_{\mu}Q^{c}&=& [\partial_{\mu}+i\frac{g_{R}}{2} \vec{\tau}\cdot\vec{W}_{\mu R}+i\frac{g_V}{6}V_{\mu}]Q^{c} \nonumber \\
D_{\mu}\Delta &=& \partial_{\mu}\Delta-i\frac{g_{L}}{2} [\vec{\tau}\cdot\vec{W}_{\mu L},\Delta]-ig_VV_{\mu}\Delta \nonumber \\
D_{\mu}\overline{\Delta} &=& \partial_{\mu}\overline{\Delta}-i\frac{g_{L}}{2} [\vec{\tau}\cdot\vec{W}_{\mu L},\overline{\Delta}]+ig_VV_{\mu}\overline{\Delta} \nonumber \\
D_{\mu}\Delta^c &=& \partial_{\mu}\Delta^c+i\frac{g_{R}}{2} [\vec{\tau}\cdot \vec{W}_{\mu R},\Delta^c]+ig_VV_{\mu}\Delta^c \nonumber \\
D_{\mu}\overline{\Delta^c} &=& \partial_{\mu}\overline{\Delta^c}+i\frac{g_{R}}{2} [\vec{\tau}\cdot\vec{W}_{\mu R},\overline{\Delta^c}]-ig_VV_{\mu}\overline{\Delta^c}~.
\end{eqnarray}
The covariant derivatives for $\widetilde\Delta$,$\widetilde{\overline{\Delta}}$,$\widetilde{\Delta^c}$,$\widetilde{\overline\Delta^c}$ have similar form as $\Delta$,${\overline{\Delta}}$,${\Delta^c}$,${\overline\Delta^c}$ respectively.

We now turn to some details of the calculation of the masses of doubly-charged Higgs boson
\cite{huitu,lrsusy2,Chacko:1997cm,dcbtheory} and the Higgsinos. This will show that these particles are indeed
light and will help us in our analysis later on.  In the context of type-II seesaw mechanism without supersymmetry,
signatures of doubly-charged Higgs bosons at the LHC has been studied in Ref. \cite{han} and in
Ref. \cite{goran} recently.
The main difference in our study is the inclusion of doubly-charged Higgsino, which helps enhance the multi-lepton signals.

\subsection{Doubly-charged Higgs boson}

The right-handed doubly-charged Higgs boson mass-squared matrix is given at tree-level as:
\begin{eqnarray}
M_{\delta^{++}}^2&=&\left( \begin{array}{cc}
-2g_R^2({\left|v_R\right|}^2 - {\left|\overline{v_R}\right|}^2)-\frac{\overline{v_R}}{v_R}Y&Y^{*} \\ Y& 2g_R^2({\left|v_R\right|}^2 - {\left|\overline{v_R}\right|}^2)-\frac{v_R}{\overline{v_R}}Y \end{array} \right)
\end{eqnarray}
where $$Y = \lambda A_{\lambda}S+|\lambda|^2(v_R\overline{v}_R-\frac{M_R^2}{\lambda})~.$$
Solving for the squared mass, it can be seen that one of the eigenvalues is negative. Including the contribution from the one-loop correction to the mass the eigenvalues become \cite{lrsusy2}
\begin{eqnarray}
M^2_{\delta^{\pm \pm}} &=& \frac{-Y (|v_R|^2 + |\overline{v_R}|^2)\pm \sqrt{(|v_R|^2 + |\overline{v_R}|^2)^2|4 g_R^2 v_R \overline{v_R}-Y|^2+4|v_R|^2|\overline{v_R}|^2|Y|^2}}{2 |v_R||\overline{v_R}|} \nonumber \\
&+&\mathcal{O} (\frac{M_{SUSY}^2}{16 {\pi}^2})
\end{eqnarray}
where $M_{SUSY}$ is the mass scale for the supersymmetry breaking which we assume to be $\sim$ 1 TeV. The factor of $1/(16 \pi^2)$ factor comes from the Coleman-Weinberg potential formula which is used to calculate the one-loop correction. Explicit calculation of the effective potential utilizing
the Majorana Yukawa couplings of the right-handed neutrino has shown that the eigenvalue which is negative at the tree-level can be made
positive \cite{lrsusy2}, thus making the symmetry breaking consistent.  This makes the mass of the right-handed doubly-charged Higgs boson to be of the electroweak scale, of order few hundred GeV. It is  naturally lighter than the doubly-charged Higgsino, since there is no loop suppression
for its mass. This light doubly-charged Higgs boson will be denoted as $\delta_R^{\pm \pm}$ in this paper.

A light doubly-charged Higgs boson can also be obtained in left-right supersymmetric models which include non-renormalizable operators in the superpotential \cite{aulakh}. Terms in the superpotential of the type $(\Delta^c \bar{\Delta}^c)^2/M_{\rm Pl}$ will give mass to the doubly-charged Higgs bosons
and Higgsinos of order few hundred GeV without resort to the Coleman-Weinberg effective potential, provided that the $SU(2)_R$ breaking scale
is in the range of $v_R \sim (10^{11}-10^{12})$ GeV.  Our analysis will also be valid for these models with light doubly-charged particles.

The left-handed doubly-charged Higgs boson mass-squared matrix looks very similar to the right-handed case except that the VEVs of the right-handed neutral Higgs boson fields are now replaced by the VEVs of the left-handed fields which we assume to be negligible. Hence the mass of the left-handed doubly-charged Higgs boson become of the order of $M_R$, which is of the scale of the $SU(2)_R$ symmetry breaking and hence large. This happens because in the Higgs boson potential, there is a cancellation between the terms $|\lambda|^2(v_R\overline{v}_R)$ and $\frac{M_R^2}{\lambda}$, arising
from the vanishing of the $F$-terms, which is not present for the left-handed doubly-charged Higgs boson mass-squared matrix. We will denote the left-handed doubly charged Higgs boson as $\delta_L^{\pm \pm}$.

\subsection{Doubly-charged Higgsino}

The right-handed doubly-charged Higgsino gets its mass only from the superpotential Eq. (\ref{sup}) and has the form $\lambda\left< S \right>$. In the supersymmetric limit, $\left< v_R \right> = \left< \overline v_R \right>$ (which arises from the vanishing of the $D$ terms)
and $\left<S\right>=0$ (which arises from the vanishing of the $F$ terms), and thus the Higgsino mass is zero in this limit.
After supersymmetry breaking, the singlet $S$ gets a VEV which is of the order of $M_{SUSY}$. Taking $\lambda$ to be of order one, we see that its mass is at the SUSY breaking scale. Thus the Higgsino has to be relatively light if we consider supersymmetry to be broken at a scale of $\sim$ 1 TeV.

The left-handed doubly-charged Higgsino would become heavy if we turn on the $W'$ term in the superpotential. In this paper we will consider $W'=0$ and hence the left-handed and the right-handed doubly-charged Higgsinos remain degenerate.  However, the case of left-handed Higgsino being heavy
can be inferred from our results, since we separate out its contribution to the four lepton plus missing $\slashed{E}_T$ final states.

\begin{center}
\begin{figure}[ht!]
\centering
	\includegraphics[width=3.2in]{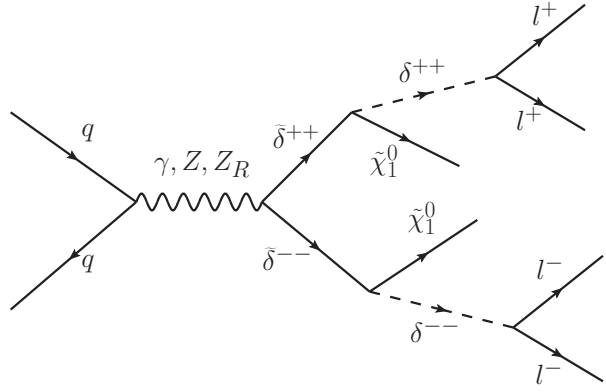}
\caption{\it
Direct production of $\widetilde{\delta}_R^{\pm\pm}$ pair at the LHC.
Subsequent decays of $\widetilde{\delta}_R^{\pm\pm}$ give rise to two leptons plus
missing energy signal, if $M_{\;\delta_R^{\pm\pm}}<M_{\widetilde{\;\delta}_R^{\pm\pm}}$.}
\label{fig:pairprod}
\end{figure}
\end{center}
\section{Signals of doubly-charged scalars and fermions at LHC} \label{sec:signals}
In this section we discuss the signal for doubly charged Higgsinos at LHC and analyze the final states coming from
the pair-production of the doubly-charged Higgsinos and their subsequent decay.\footnote{The relevant Feynman rules are listed in the Appendix.} The doubly charged Higgsinos are pair-produced at the LHC through the process
$$p\, p \longrightarrow \widetilde{\delta}_{L,R}^{++}  \widetilde{\delta}_{L,R}^{--}~~~~{\rm (illustrated ~in ~Fig.~\ref{fig:pairprod})}$$
which proceeds through $s$-channel $\gamma$ and $Z_{L,R}$ exchanges \cite{rai:2008}. As the mass
of $Z_R$ is dependent on the scale at which the $SU(2)_R$ is broken, its contribution will vary depending upon its allowed values. In the minimal left-right supersymmetric model, there is a relation between the $W_R$ and the $Z_R$ mass where $M_{Z_R} \sim 1.7 M_{W_R}$. Therefore the current limit on the $W_R$ mass of about 2.5 TeV \cite{WRsearch} requires the $Z_R$ to be rather heavy. This heavy $Z_R$ has very small contributions to the pair-production cross section of the doubly charged Higgsinos. In our analysis we have fixed the $Z_R$ mass at 5 TeV and find that the contributions from $Z_R$ exchange only become comparable to the electroweak gauge boson
exchanges for large values of the doubly charged Higgsino mass, where the overall signal is quite suppressed.

We focus on a natural scenario where the only ``light" states beyond the SM are the doubly-charged Higgs boson, doubly-charged Higgsino and the lightest neutralino, which is the LSP.  The left-handed doubly-charged Higgsino is degenerate with the right-handed doubly-charged Higgsino
(in the case where $W' = 0$).  All other SUSY particles are assumed to be much heavier.
We further assume that the doubly-charged Higgsino is heavier than the right-handed doubly charged Higgs boson and the lightest neutralino.
Then the dominant decay channel for the doubly-charged Higgsino is to the light doubly-charged Higgs boson and the LSP neutralino,
which we assume is allowed by kinematics.  The branching ratio for this process is
almost 1 in this scenario as the next leading decay mode is into a lepton and an off-shell slepton
which is highly suppressed. The right-handed doubly-charged Higgs boson now decays almost entirely into two same sign leptons giving rise to a final signal of 4 leptons and missing energy. Other decay modes of the right-handed doubly-charged Higgs boson would be into two real or virtual
$W_R$ bosons or a $W_R$ and a single-charged Higgs boson. Both the $W_R$ and the single-charged Higgs boson are very heavy in this model and hence those decays will be forbidden or highly suppressed. The entire decay chain is then,
\bi
\item $ \widetilde{\delta}_R^{\pm\pm} \rightarrow \delta_R^{\pm\pm} \widetilde\chi_1^{0} $
\item$\delta_R^{\pm\pm} \rightarrow \ell^{\pm} \ell^{\pm} $
\ei

Though the right-handed doubly-charged Higgsino decays almost always into a right-handed doubly-charged Higgs boson and a neutralino, the left-handed doubly-charged Higgsino which is degenerate with the right-handed doubly-charged Higgsino cannot decay through this channel as the left-handed doubly-charged Higgs boson is much heavier. The main decay channel for the left-handed Higgsino is then given by the three-body decay  through an off-shell slepton and a lepton, where the off-shell slepton mediates the decay into a lepton and a neutralino \cite{rai:2008}. This produces the same final state product as our signal and is therefore a source of background if we consider the signal coming only from the right-handed
doubly charged Higgsinos. The left-handed doubly-charged Higgsino production cross-section is larger than the right-handed Higgsino due to the $Z$-boson coupling strength being larger to the left-handed particles and hence we also need to analyze the decay of the left-handed Higgsino and include its contributions. We must however note that both
the right-handed and left-handed Higgsino pair production leads to a four-lepton final state
with large missing transverse momenta because of the presence of the undetected LSP passing
through the detector. Another source for the four-lepton final state would come from the
pair production of the light doubly-charged Higgs boson present in the model. Presence of such
doubly-charged Higgs bosons have been looked for by experimentalists in the context of various
other models at Tevatron as well as LHC \cite{dcbexperiment} which put strong limits on the
masses of such particles.

\begin{figure}[b!]
\centering
	\includegraphics[width=3.5in]{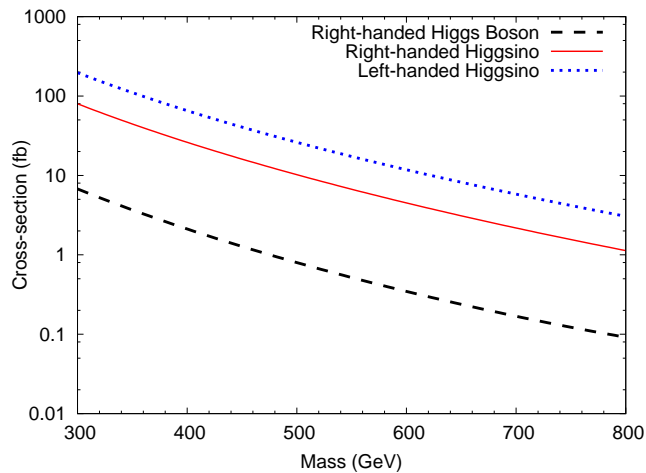}
\caption{\it
Production cross sections for $\widetilde{\delta}_{L,R}^{\pm\pm}$ pair and $\delta_R^{\pm\pm}$ at the LHC at 14 TeV}.
\label{fig:prodcomp}
\end{figure}
%
In Fig.~\ref{fig:prodcomp} we plot the production cross-sections for the pair production of doubly-charged Higgsinos (both chirality) as well
as for the right-handed doubly-charged Higgs boson. Note that the production cross section
for the left-handed doubly-charged Higgsino is much larger than the right-handed one.
This is due to the bigger $Z$ boson coupling with the left-handed doubly-charged Higgsino. However for larger values of the mass, the required center of mass energy
to produce the particles in pair also increases and therefore an $s$-channel suppression
would appear in the case of the left-handed doubly-charged Higgsino as the center of mass energy moves away from the $Z$ boson pole mass,  i.e.
$\frac{1}{\hat{s}-M_Z^2} \to \frac{1}{\hat{s}}~ (\hat{s}>>M_Z^2)$. In comparison the $Z_R$
contribution would increase as the center of mass energy starts approaching the $Z_R$
boson pole mass, i.e. $\frac{1}{\hat{s}}\to\frac{1}{\hat{s}-M_{Z_R}^2}~ (\hat{s} \sim M_{Z_R}^2)$
which also has larger coupling to the right-handed doubly-charged Higgsino. This effect is visible for very
large values of the Higgsino mass (although not shown in the Fig. \ref{fig:prodcomp}) where we find that the production cross section for the left-handed Higgsino actually falls below the production cross section of
the right-handed Higgsino. It can also be seen that the Higgsino production cross-sections are much larger than the doubly-charged Higgs boson production rate (for the same mass) and hence they effectively help in enhancing
the 4-lepton signal at colliders.  In general, from spin arguments we might expect the production
cross-section of the fermion to be four times that of the scalar, but this is only true in the massless limit. One can think that since the center of mass energy is much higher than the masses of the particles the massless limit should be a good approximation, but turning on the parton distribution function produces partons of
all energies and hence we get a cross-section ratio which is much higher. The Higgsino process also
gives a unique signal with $4 \ell + \slashed{E}_{T}$ which is not present for the doubly-charged Higgs
boson pair-production process.

Considering the decays of the doubly-charged particles discussed before, we find that the final states coming from the pair production and subsequent decays of the doubly-charged Higgsinos are two pairs of same-sign leptons of same or different flavor (i.e., $e$ or  $\mu$) and missing energy.
We want to focus on all the possibilities with the final states consisting of same flavor or different flavor leptons, with and without missing energy.

As we have no hint of SUSY signals yet at the LHC, it can be safely assumed that the SUSY particles are
heavy and difficult to produce at the current energies at which LHC was run. We therefore restrict ourselves
to the low lying mass spectrum of some of the SUSY particles and their decay
probabilities to study its signals. Since the model in study naturally accommodates light doubly-charged particles, we assume all other SUSY
partners as well as the Higgs scalars to be much heavier than the doubly-charged Higgsinos and the doubly charged Higgs boson
(from the right-handed sector). The only other particle which is assumed to be lighter is the lightest neutralino, which is the LSP. With this choice of the spectrum, the decay patterns for the doubly charged particles are known and have already been discussed earlier.
To highlight the signal we have considered two representative points :
\bi
\item The first choice, which we call {\bf BP1} (Benchmark Point 1), we consider a doubly-charged Higgs boson with mass 300 GeV, an LSP neutralino with a mass of 80 GeV, charged sleptons with mass of 1 TeV and doubly-charged Higgsinos with a mass of 500 GeV. With this choice we focus our attention on two particular scenarios.
First, we analyze the situation where all the finals state leptons coming from the decay are of the same
flavor (e.g all the final state leptons are either electrons or muons) while the other case is when each
doubly-charged particle decays to a different flavor pair (e.g. two same sign electrons and two same sign muons).
\item The second choice, which we call {\bf BP2},  we consider a lower value for the mass of doubly-charged Higgsino as 400 GeV while the
other mass choices remain the same. Note that this choice gives a larger production rate for the doubly-charged Higgsinos, but also affects
the kinematics of the final state decay products because of smaller mass splitting between the doubly-charged Higgsino and the doubly-charged Higgs boson.
\ei

In our analysis, for the charged lepton final states we have considered the signal consisting of either electrons or muons only and
neglected the tau lepton.  Nevertheless the decay of the doubly-charged Higgs boson to tau lepton pair will be very similar to the decay into muons and electrons and is only considered less relevant due to the limited tau-tagging efficiency at experiments. However, the signal will
also be dictated by the decay probabilities of the doubly-charged scalar into the charged lepton pairs, and in models where  the Yukawa structure demands that the decays are maximally to a pair of same sign taus, then one needs to consider the tau final states.

We now turn our focus to analyzing the final state signal consisting of the four charged leptons with or without missing transverse energy.
Note that when we do not demand any criterion for the missing transverse momenta in the final state, our signal contributions come from
three different sources, {\it i.e.} pair production of the doubly-charged Higgsinos (both chirality) as well as the pair production of the
doubly-charged scalars. This would not only enhance the four-lepton signal when compared to individual contributions but also help in
identifying the nature of additional contributions to such multi-lepton final states. To study the signal we demand that the final state particles
satisfy the following kinematic cuts:
\bi
\item Each charged lepton must carry a minimum transverse momentum given by
$p_T >$ 15 GeV.
\item The charged leptons must lie in the central rapidity region of $ |\eta_{\ell}|<2.5$.
\item For proper resolution to detect the final state particles we set $\Delta R_{\ell\ell}>0.2$ between the final state charged leptons, where $\Delta R = \sqrt{\left( \Delta \phi\right) ^2+\left( \Delta \eta\right) ^2}$ defines the resolution of a pair of particles in the $(\eta,\phi)$ plane.
\item We also specify an invariant mass cut between the opposite sign same flavor leptons
such that $M_{\ell^+\ell^-}>10\ GeV$ and a further cut of $80\ GeV>M_{\ell^+\ell^-}>100\ GeV$, where the latter one is aimed at removing the SM contributions coming from resonant $Z$ boson decays.
\ei

With the above set of kinematic selections we perform a detailed numerical analysis of the
final state events of the multilepton signal as well as the SM background. For our
numerical analysis, we have included the
model description into the event generator {\tt  CalcHEP} \cite{calchep}
and generated the event files for the production and decays of the doubly-charged Higgsinos. These event files were then passed through the {\tt CalcHEP+Pythia} \cite{pythia} interface where we include the effects of both initial and final state radiations using Pythia switches to smear the final states. We have used the leading order
CTEQ6L \cite{cteq} parton distribution functions (PDF) for our analysis.

So there are three major processes that contribute to out signal.
\bi
\item The direct pair-production of the right-handed doubly-charged Higgs bosons. Each Higgs boson then decays into a pair of same sign leptons producing a final state signal of 4 leptons. We call this $({\mathcal C1})$
$$p~p \rightarrow \delta_R^{++} \delta_R^{--}\rightarrow \ell^{+}_i \ell^{+}_i\ell^{-}_j\ell^{-}_j $$
\item Pair-production of right-handed doubly-charged Higgsino. Each Higgsino decays into a right-handed doubly-charged Higgs boson and a neutralino. The doubly-charged Higgs boson then decays into a pair of same-sign leptons giving a final state signal of 4 leptons and $\slashed{E}_T$. We call this $({\mathcal C2})$
$$p~p \rightarrow \widetilde\delta_R^{++} \widetilde\delta_R^{--}\rightarrow \delta_R^{++} \delta_R^{--} \widetilde\chi_1^0\widetilde\chi_1^0\rightarrow \ell^{+}_i \ell^{+}_i\ell^{-}_j\ell^{-}_j \slashed{E}_T$$
\item Pair-production of left-handed doubly-charged Higgsino. The Higgsino decays
through an off-shell slepton to
a same sign lepton pair and a neutralino. This process also gives a final state signal with 4 leptons and $\slashed{E}_T$. We call this $({\mathcal C3})$
$$p~p \rightarrow \widetilde\delta_L^{++} \widetilde\delta_L^{--}\rightarrow (\widetilde{\ell}^{*+}_i \ell^{+}_i)~ (\widetilde{\ell}^{*-}_j\ell^{-}_j) \rightarrow \ell^{+}_i \ell^{+}_i\ell^{-}_j\ell^{-}_j\widetilde\chi_1^0\widetilde\chi_1^0 $$
\ei
All the three subprocesses mentioned above lead to a signal with four charged leptons in the final state which
is a very clean signal at a hadron machine such as the LHC, with very little SM background,
and therefore should be an interesting test for the model. Significantly one should note that
the signal described by (${\mathcal C1}$) is an important channel for the search of doubly charged
particle resonances such as double charged scalars \cite{dcbexperiment} or bileptons \cite{bileptons} and can
appear even in R-parity violating supersymmetric models \cite{rparity}. The highlight of course is that
there is no source for missing transverse momenta in the signal. However, the other two signals
described by (${\mathcal C2}$) and (${\mathcal C3}$) not only lead to four charged leptons in the final
states but is also accompanied by large missing transverse momenta due to the LSP present in the
final state. There could be numerous new physics scenarios where such a signal can be
common and so it would be interesting to be able to identify the signal associated with our model in a
unique way.
\begin{table}[h!]
\centering
    \begin{tabular}{|c|c|c|c|c|c|c|}
        \hline
        {\bf{LHC Energy}} &\multicolumn{2}{c|} {${\mathcal C1}$} &\multicolumn{2}{c|} {${\mathcal C2}$}&\multicolumn{2}{c|} {${\mathcal C3}$}  \\
        ~ &\multicolumn{2}{c|} {$\slashed{E}_T$ (GeV)} &\multicolumn{2}{c|} {$\slashed{E}_T$ (GeV)}&\multicolumn{2}{c|} {$\slashed{E}_T$ (GeV)}  \\
        ~ &$>$ 0 & $>$ 100 & $>$ 0 & $>$ 100 & $>$ 0 & $>$ 100  \\ \hline
       7 TeV & 0.266 fb & 0.033 fb & 0.275 fb & 0.226 fb & 0.642 fb & 0.568 fb \\ \hline
        8 TeV & 0.368 fb & 0.048 fb & 0.430 fb & 0.359 fb & 0.992 fb & 0.927 fb \\ \hline
        14 TeV & 1.153 fb & 0.228 fb & 1.859 fb & 1.649 fb & 4.208 fb & 3.667 fb \\
        \hline
    \end{tabular}
    \caption{{\sl Cross-section table for a final state of~ $\ell_i^+ \ell_i^+ \ell_i^- \ell_i^- + X$ with $M_{\widetilde\delta_{L,R}^{\pm\pm}}$= 500 GeV,$M_{\delta_R^{\pm\pm}}$= 300 GeV, $M_{\widetilde\chi_1^0}$= 80 GeV and $M_{\widetilde{l}^\pm}$= 1 TeV}}
\label{tab:cs1}
\end{table}
We find that our signal can in general be classified into two types, one where we only demand
four charged leptons in the final state and do not put any requirement on the missing transverse momenta.
The other type would be to demand a minimum missing transverse momenta in the final state in addition to
the four tagged charged leptons. We list the cross-sections for the three subprocesses (${\mathcal C1}$--${\mathcal C3}$) at different LHC energies in Table \ref{tab:cs1} which gives the cross section
for a final state consisting of same-sign pairs and all four of same-flavor (SF) charged leptons in our
model for {\bf BP1} where the doubly-charged Higgsino mass is taken as 500 GeV, doubly-charged Higgs boson mass of 300 GeV, slepton mass of 1 TeV and a neutralino mass of 80 GeV. Note that the signal cross sections are invariably larger
for the $({\mathcal C3})$ as it comes from the pair production of the left-handed doubly charged Higgsinos
which has the greater production rate. We can see that without any missing $E_T$ requirement on the final
state, a somewhat lower cross section for the signal coming from the pair production of doubly charged scalar
is found to be enhanced considerably by including contributions from the pair production of the doubly charged Higgsinos. This enhances the sensitivity of the experiment to exotic doubly charged particles through the
four charged lepton final state. With a minimum missing $E_T$ requirement of 100 GeV on the events, it is
found that the signal coming from the pair production of the doubly charged scalars is reduced drastically
while the events from the pair production of the doubly charged Higgsinos are not affected much. This is expected
because the doubly charged Higgsinos decay to final states consisting of the undetected LSP which carries off
substantial missing energy and therefore satisfies the large $\slashed{E}_T$ cut-off.
\begin{table}[h!]
\centering
    \begin{tabular}{|c|c|c|c|c|c|c|}
        \hline
        {\bf{LHC Energy}} &\multicolumn{2}{c|} {${\mathcal C1}$} &\multicolumn{2}{c|} {${\mathcal C2}$}&\multicolumn{2}{c|} {${\mathcal C3}$}  \\
        ~ &\multicolumn{2}{c|} {$\slashed{E}_T$ (GeV)} &\multicolumn{2}{c|} {$\slashed{E}_T$ (GeV)}&\multicolumn{2}{c|} {$\slashed{E}_T$ (GeV)}  \\
          ~ &$>$ 0 & $>$ 100 & $>$ 0 & $>$ 100 & $>$ 0 & $>$ 100  \\ \hline
        7 TeV & 0.302 fb & 0.032 fb & 0.314 fb & 0.257 fb & 0.753 fb & 0.672 fb \\ \hline
        8 TeV & 0.418 fb & 0.047 fb & 0.480 fb & 0.402 fb & 1.152 fb & 1.078 fb \\ \hline
        14 TeV & 1.266 fb & 0.216 fb & 1.989 fb & 1.749 fb & 4.655 fb & 4.051 fb  \\
        \hline
    \end{tabular}
    \caption{{\sl Cross-section table for a final state of~ $\ell_i^+ \ell_i^+ \ell_j^- \ell_j^- + X$ with $M_{\widetilde\delta_{L,R}^{\pm\pm}}$= 500 GeV,$M_{\delta_R^{\pm\pm}}$= 300 GeV, $M_{\widetilde\chi_1^0}$= 80 GeV and $M_{\widetilde{l}^\pm}$= 1 TeV}}
\label{tab:cs2}
\end{table}
In Table \ref{tab:cs2} we show the cross-section for a final state consisting of same-sign pairs where each pair is of different-flavor (DF) leptons for {\bf BP1}. Here we assume that one of the doubly-charged
particle decays to one particular flavor of the charged leptons while the other decays to a different flavor.
So the final states would have four charged leptons of the type $e^\pm e^\pm \mu^\mp \mu^\mp$.  Note
that such a combination of final state would have practically no SM background as it requires at least four
$W$ bosons to give such a combination of charged leptons in the final state. We neglect the $\tau$ lepton
as discussed before. The cross sections are slightly greater than those listed in Table \ref{tab:cs1} because
we have removed the additional kinematic cut on the invariant mass on the opposite-sign same flavor leptons given by $80\ GeV>M_{\ell^+\ell^-}>100\ GeV$. As our estimates rely on the assumption that the branching
fractions for the doubly charged particles decay to each flavor of charged lepton is 1/3, we must point out
that this final state will be relevant only when the decay rates to either $e^\pm e^\pm$ or $\mu^\pm \mu^\pm$ are not too suppressed.
\begin{figure}[h!]
\includegraphics[width=2.9in]{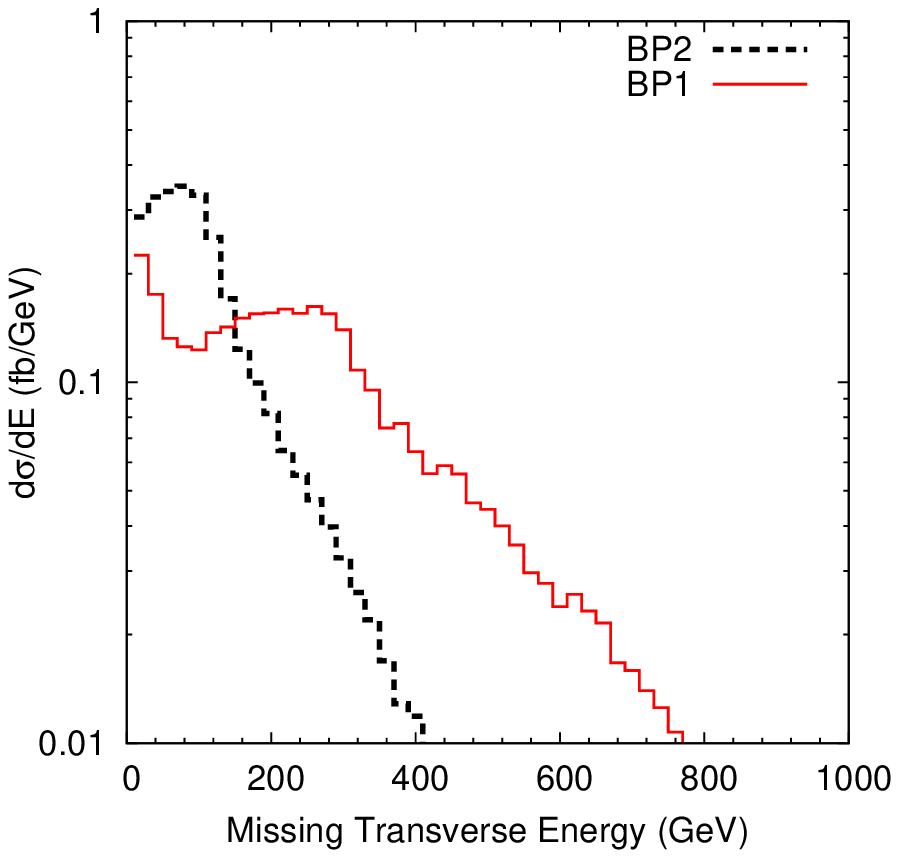}
\includegraphics[width=2.9in]{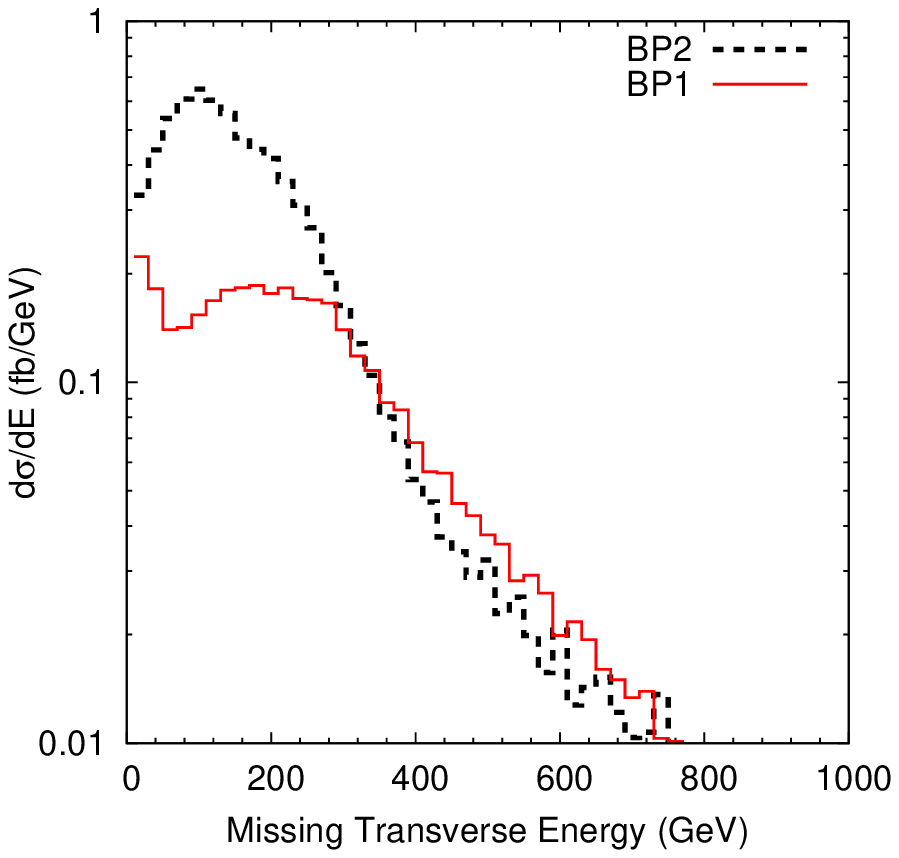}
\caption{{\it (a) $\slashed{E}_T$ for doubly-charged Right-handed Higgsino and Higgs boson, (b) $\slashed{E}_T$ for doubly-charged Right-handed and Left-handed Higgsinos and Right-handed Higgs boson.}}
\label{fig:met}
\end{figure}

We now consider a case where the doubly charged Higgsinos are slightly lighter ($400$ GeV) while the
other particles have the same mass as before ({\bf BP2}). This choice enhances the production rates for
the doubly-charged Higgsinos but also gives a compressed spectrum for its decays. Note that a bigger mass difference
between the parent particle and its decay products would lead to greater energy for the decay products.
In this case, one expects that as the LSP mass and the doubly charged Higgs mass add up very close to
the doubly-charged Higgsino mass, the missing transverse momenta in the events due to the LSP will
be less compared to the previous case. This can be seen in Fig. \ref{fig:met} where we show the
distribution for the differential cross section as a function of the missing transverse energy.
In Fig.~\ref{fig:met}(a) we show the $\slashed{E}_T$ distribution in the signal events coming from the contributions of the right-handed doubly-charged Higgsino and Higgs while Fig.~\ref{fig:met}(b) shows $\slashed{E}_T$ distribution for contributions from both the right-handed and left-handed doubly-charged Higgsino including the doubly charged Higgs boson. We see that differential cross section in
Fig. \ref{fig:met}(a) has a higher fraction of events at very small $\slashed{E}_T$. This is because of the contribution from the direct pair production of the doubly-charged Higgs boson which will have very little missing energy which might originate due to mismeasurements of the final state particles, as there is no other source of missing energy in the form of the
neutralino in the final state. In Fig.~\ref{fig:met}(b) this effect is washed away because the
\begin{table}[h!]
    \begin{tabular}{|c|c|c|c|c|c|c|}
        \hline
        {\bf{LHC Energy}} &\multicolumn{2}{c|} {${\mathcal C1}$} &\multicolumn{2}{c|} {${\mathcal C2}$}&\multicolumn{2}{c|} {${\mathcal C3}$}  \\
        ~ &\multicolumn{2}{c|} {$\slashed{E}_T$ (GeV)} &\multicolumn{2}{c|} {$\slashed{E}_T$ (GeV)}&\multicolumn{2}{c|} {$\slashed{E}_T$ (GeV)}  \\
          ~ &$>$ 0 & $>$ 20 & $>$ 0 & $>$ 20 & $>$ 0 & $>$ 20  \\ \hline
        7 TeV & 0.266 fb & 0.143 fb & 0.871 fb & 0.823 fb & 1.797 fb & 1.774 fb  \\ \hline
        8 TeV & 0.368 fb & 0.203 fb & 1.248 fb & 1.183 fb & 2.576 fb & 2.550 fb  \\ \hline
        14 TeV & 1.153 fb & 0.737 fb & 4.467 fb & 4.309 fb & 8.892 fb & 8.806 fb  \\
        \hline
    \end{tabular}
    \caption{{\sl Cross-section table for a final state of~ $\ell_i^+ \ell_i^+ \ell_i^- \ell_i^- + X$ with $M_{\widetilde\delta_{L,R}^{\pm\pm}}$= 400 GeV,$M_{\delta_R^{\pm\pm}}$= 300 GeV, $M_{\widetilde\chi_1^0}$= 80 GeV and $M_{\widetilde{l}^\pm}$= 1 TeV}}
\label{tab:cs3}
\end{table}
number of events from the left handed doubly-charged Higgsino pair-production is now much larger compared to both the doubly-charged Higgs boson and Higgsino pair-production and hence their contribution is suppressed.

In Table \ref{tab:cs3} we give the cross sections for a final state consisting of the same-flavored charged leptons for {\bf BP2}. Note that we have a slightly weaker requirement on the missing transverse energy
of 20 GeV for the events corresponding to {\bf BP2}. This is to avoid large suppression of the signal
which can happen due to the smaller mass splittings.
\begin{table}[h!]
\centering
    \begin{tabular}{|c|c|c|c|c|c|c|}
        \hline
        {\bf{LHC Energy}} &\multicolumn{2}{c|} {${\mathcal C1}$} &\multicolumn{2}{c|} {${\mathcal C2}$}&\multicolumn{2}{c|} {${\mathcal C3}$}  \\
        ~ &\multicolumn{2}{c|} {$\slashed{E}_T$ (GeV)} &\multicolumn{2}{c|} {$\slashed{E}_T$ (GeV)}&\multicolumn{2}{c|} {$\slashed{E}_T$ (GeV)}  \\
          ~ &$>$ 0 & $>$ 20 & $>$ 0 & $>$ 20 & $>$ 0 & $>$ 20  \\ \hline
         7 TeV & 0.302 fb & 0.149 fb & 1.009 fb & 0.949 fb & 2.332 fb & 2.308 fb  \\ \hline
        8 TeV & 0.418 fb & 0.213 fb & 1.451 fb & 1.358 fb & 3.327 fb & 3.288 fb  \\ \hline
        14 TeV & 1.266 fb & 0.721 fb & 4.804 fb & 4.610 fb & 10.886 fb & 10.767 fb \\
        \hline
    \end{tabular}
    \caption{{\sl Cross-section table for a final state of~ $\ell_i^+ \ell_i^+ \ell_j^- \ell_j^- + X$ with $M_{\widetilde\delta_{L,R}^{\pm\pm}}$= 400 GeV,$M_{\delta_R^{\pm\pm}}$= 300 GeV, $M_{\widetilde\chi_1^0}$= 80 GeV and $M_{\widetilde{l}^\pm}$= 1 TeV}}
\label{tab:cs4}
\end{table}

In Table \ref{tab:cs4} we give the cross sections for a final state consisting of different-flavored
charged lepton pairs for {\bf BP2}. Again the kinematic characteristics for the events remain the same
as before but the cross section is slightly greater than that for SF events because of the removal of the kinematic cut corresponding to the invariant mass removing the $Z$ peak for opposite sign same
flavor charged lepton pairs.

We must point out here that the corresponding SM background for the four charged lepton final state with
our selection cuts on the kinematic variables is found to be completely negligible and therefore has not
been shown or considered in our analysis. The most dominant background which one expects for the
SF charged lepton signal will be from the pair production of $Z$ bosons which we have suppressed using the
invariant mass cut on the opposite-sign same flavor lepton pairs. However, as we have a light
doubly-charged Higgs in the spectrum, we expect to see a resonance in the invariant mass distributions
of like-signed charge lepton pairs. We have already shown that there are three different subprocesses
for the signal contributions for the $4\ell + X$ final state and the cross-section for $({\mathcal C3})$ is
much larger than $({\mathcal C1})$ and $({\mathcal C2})$. Note that $({\mathcal C3})$ corresponds to the signal where the left-handed doubly-charged Higgsino is pair produced and decays through an off-shell slepton.
Therefore one does not expect any resonance behavior in the invariant mass distributions of the charged lepton pairs but a kinematic edge is expected \cite{rai:2008}. This would mean that a part of the signal itself acts as a background to smear out the resonant signal for the doubly
charged Higgs boson. This is in fact the highlight of our analysis where we show that
our signal actually stands out as a resonance and is also enhanced because of the additional contributions
coming from the heavy doubly charged fermion production.

To show some kinematic characteristics of the events for the SF signal we take the case of
$e^+e^+e^-e^-$ in the final state and for the DF signal we take $\mu^-\mu^-e^+e^+$. We put the aforementioned cuts and simulate the events using CalcHEP and Pythia and look at the $\Delta{R}_{ll}$
and invariant mass $M_{ll}$ of the final state leptons.

\begin{figure}[ht!]
\includegraphics[width=3.1in]{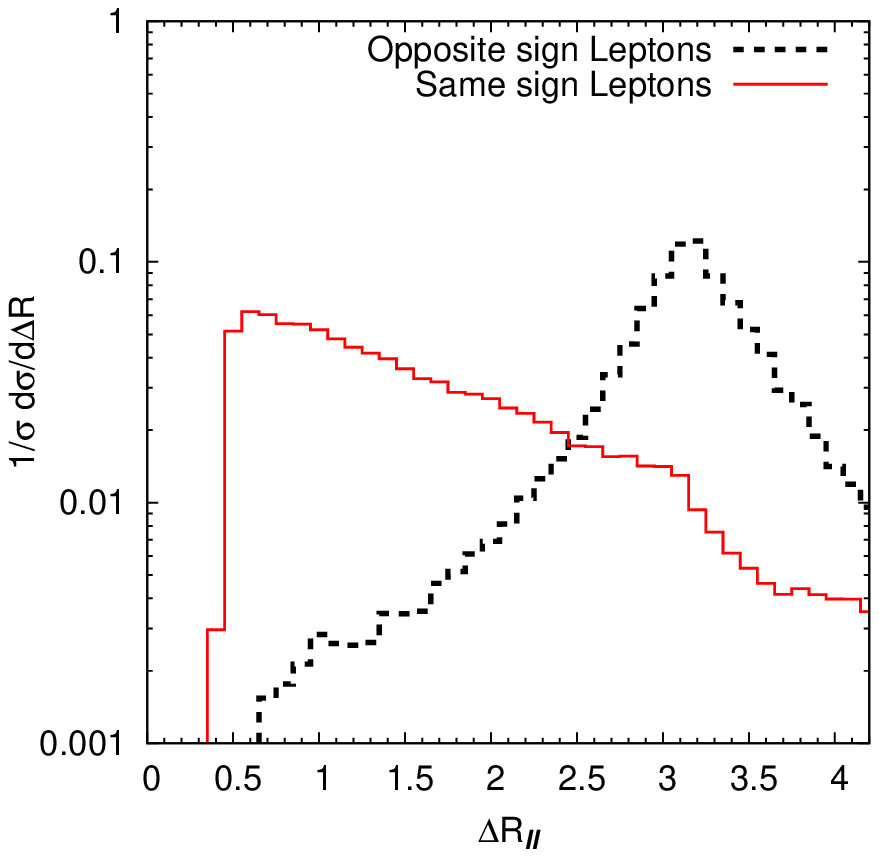}
\includegraphics[width=3.1in]{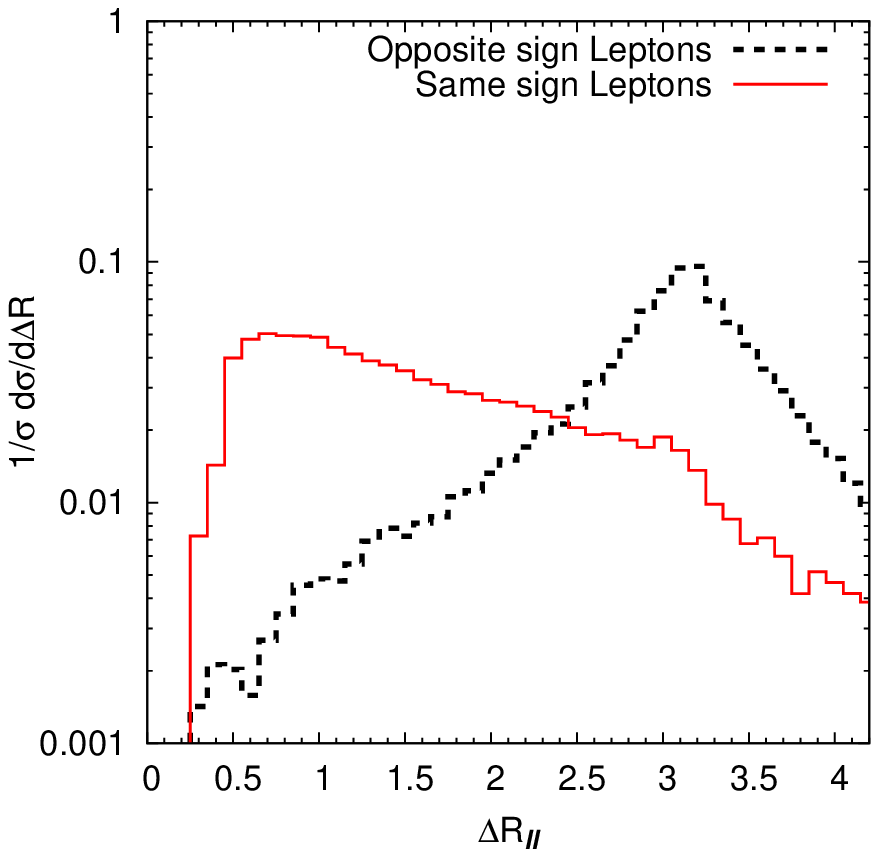}
\caption{{\it (a) Illustrating the $\Delta{R}_{ll}$ distribution for events coming from the doubly-charged right-handed Higgsino and Higgs boson pair production and, (b) $\Delta{R}_{ll}$ distribution for events
when the contributions from the pair production of the left-handed Higgsinos is also included for {\bf BP1}.}}
\label{fig:drssee}
\end{figure}

The $\Delta{R}_{ll}$ for the same-sign and opposite-sign final state charged leptons of same
flavor for {\bf BP1} are shown in Fig. \ref{fig:drssee}. Fig. \ref{fig:drssee}(a) includes only the contribution of the right-handed doubly-charged Higgsino and Higgs (${\mathcal C1+C2}$) while
Fig. \ref{fig:drssee}(b) denotes the contribution from the doubly-charged Higgs as well as both the right-handed and left-handed doubly-charged Higgsino (${\mathcal C1+C2+C3}$). It is worth noting that in each plot there
is a marked difference  between the same-sign lepton and the opposite-sign leptonic final states. It can be seen that for the same-sign charged leptons the distribution is peaked at low values of $\Delta{R}$ while the opposite-sign charged leptons have a $\Delta{R}$ which is peaked at a much higher value. This is what is expected since the same-sign pair of leptons arise from the decay of a single doubly-charged Higgs boson while the opposite-sign leptons arise from two different particles and hence are much further apart. The measurement of $\Delta{R}$ at the LHC for a four lepton final state can thus give a definite indication of the existence of a doubly-charged particle if the distribution is similar to what we get in our analysis. Note that the
$\Delta R$ distributions are also very sensitive to the boost of the mother particle as larger
boost will make the decay products come out more closer to each other.

\begin{figure}[h!]\centering
\includegraphics[width=3.1in]{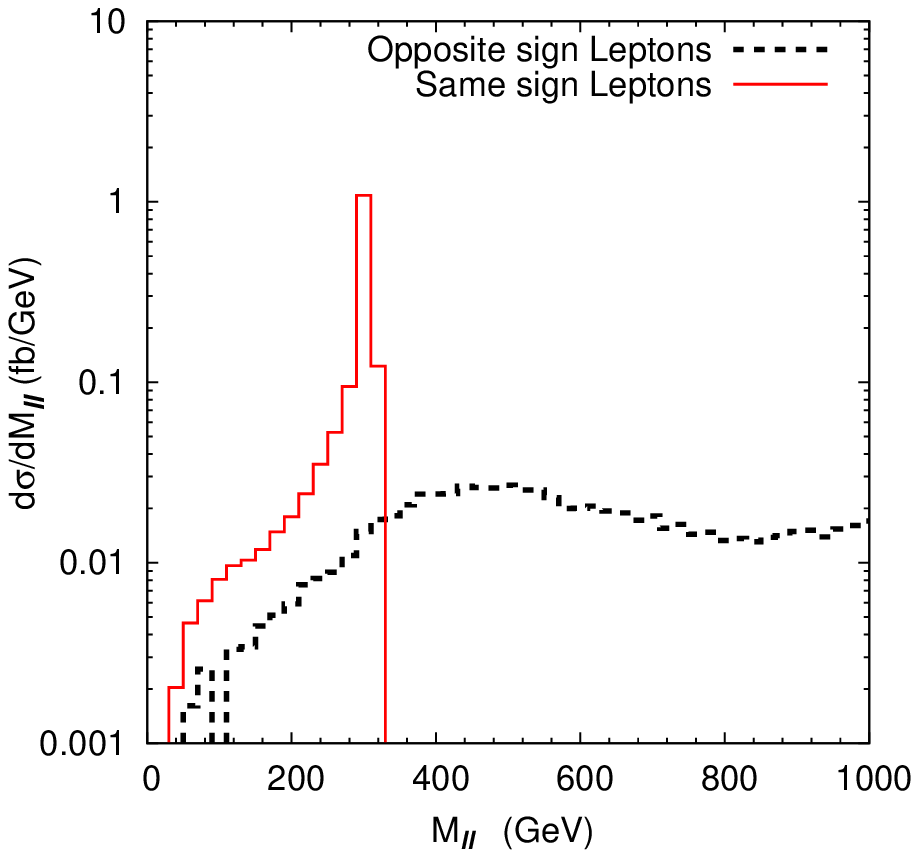}
\includegraphics[width=3.1in]{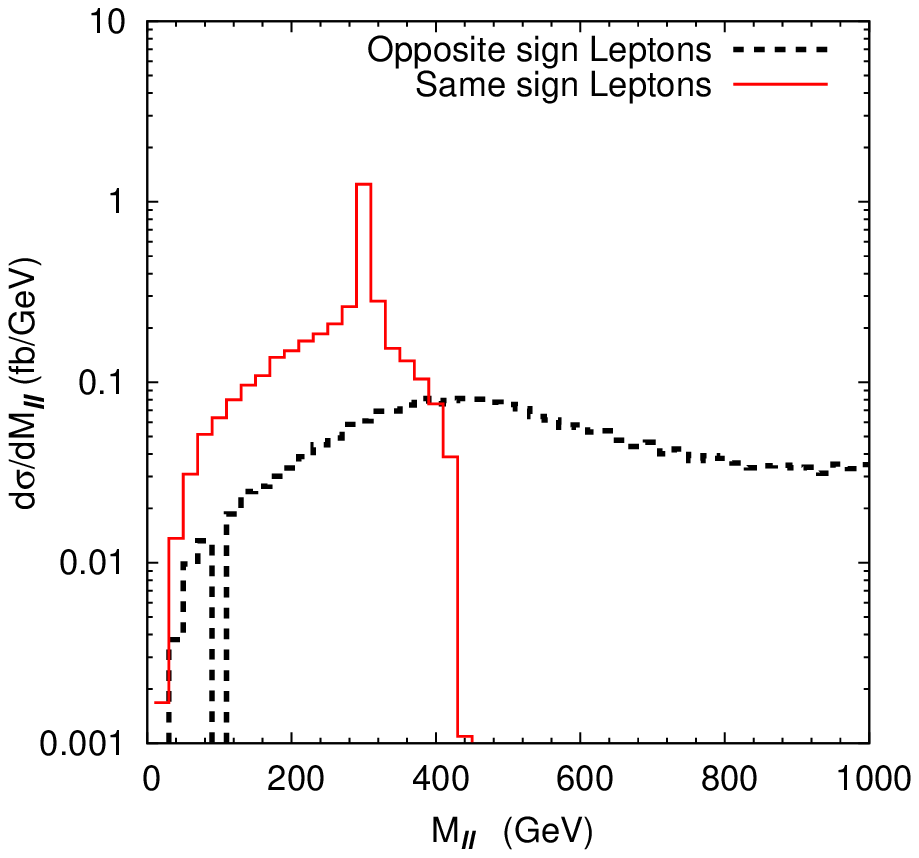}
\caption{{\it Illustrating the (a) invariant mass distribution for events coming from the doubly-charged right-handed Higgsino and Higgs boson pair production and,
(b) invariant mass distribution for events
when the contributions from the pair production of the left-handed Higgsinos is also included for {\bf BP1}.}}
\label{fig:mosee}
\end{figure}
In Fig. \ref{fig:mosee} we show the invariant mass distributions for the same-sign and opposite-sign final state leptons of same flavor for {\bf BP1}. Note that for the opposite-sign lepton pair invariant mass there are no events between 80 GeV and 100 GeV. This is due to the cut that we applied to get rid of the $Z$ peak for the SM background. The invariant mass for the opposite-sign leptons do not show any resonant behavior.  For the
same-sign lepton pairs, we see a pronounced peak at an invariant mass of 300 GeV which is the doubly-charged Higgs boson mass. As we include the contributions coming from the pair production of the left-handed
doubly-charged Higgsino, the resonant peak is seen to broaden a little but is still very significant. Such a peak, though very difficult to see without a priori knowledge of the Higgs boson mass, would be a definite proof of a doubly-charged particle if seen in the detector.  It is also worth noting the distinct kinematic edge seen in the
invariant mass distribution of the like-sign charged lepton pair in both Fig. \ref{fig:mosee}(a) and (b). The
edge in Fig. \ref{fig:mosee}(a) is at a different $M_{ll}$ when compared to that in Fig. \ref{fig:mosee}(b).
Note that in Fig. \ref{fig:mosee}(a) the resonant peak is because of the doubly-charged Higgs decaying
to two same-sign leptons while the sharp cut-off in the distribution is because of the maximum invariant mass
allowed for the lepton pair that comes from $\delta_R^{\pm\pm} \rightarrow \ell^\pm \ell^\pm$. This would mean
that the distribution will fall rapidly beyond the resonance which is the $\delta_R^{\pm\pm}$ mass.
On the other hand, the signal in Fig. \ref{fig:mosee}(b) is completely dominated by the contributions
coming from the left-handed doubly-charged Higgsino production and therefore it washes away the
kinematic edge from the other subprocesses. The sharp cut-off in Fig. \ref{fig:mosee}(b) then appears
because of $  \widetilde\delta_L^{\pm} \rightarrow (\widetilde{\ell}^{*\pm}_i \ell^{\pm}_i)
\rightarrow \ell^{\pm}_i \ell^{\pm}_i \widetilde\chi_1^0 $ and is given by
(in the rest frame of the decaying particle)
$M_{l^\pm l^\pm}^{max} = \sqrt{M_{\widetilde{\delta}_L^{\pm\pm}}^2 + M_{\widetilde\chi_1^{0}}^2
- 2 M_{\widetilde{\delta}_L^{\pm\pm}} E_{\widetilde\chi_1^{0}}}$,
where $E_{\widetilde\chi_1^{0}}$ is the energy of the LSP. This yields an edge in the invariant mass distribution
of the same-sign same flavor charged lepton pairs at the bin around $M_{l^\pm l^\pm} = M_{\widetilde{\delta}_L^{\pm\pm}} - M_{\widetilde\chi_1^{0}} \simeq 420$ GeV. It is interesting to observe that
we find a distinct resonance in the invariant mass distribution as well as a sharp kinematic edge due to the
off-shell decay of the left-handed doubly-charged Higgsino which clearly highlights an additional contribution
to the resonant signal of doubly-charged scalar production leading to four charged lepton final states.

\begin{figure}[h!]
\includegraphics[width=3.1in]{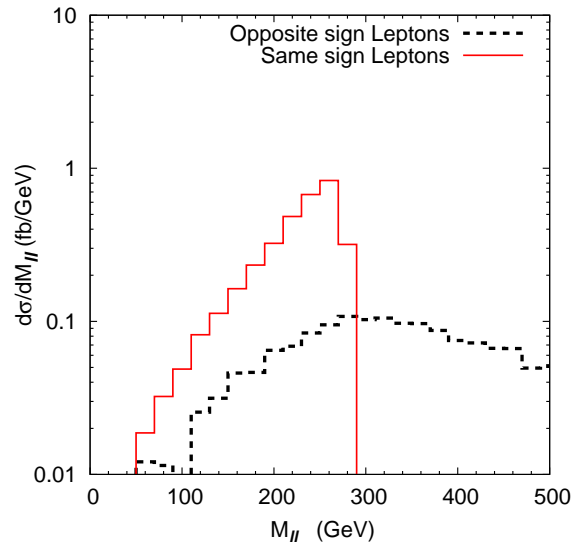}
\caption{{\it  Invariant mass distribution in $M_{ll}$ for a doubly-charged right-handed Higgsino which decays  through an off-shell doubly-charged Higgs boson.}}
\label{fig:minv}
\end{figure}
We can also consider the case where the right-handed doubly-charged Higgsino too decays via off-shell doubly charged scalar which can be realized when the right-handed doubly-charged Higgsino is not much heavier than
the right-handed doubly-charged Higgs boson such that $M_{\widetilde{\delta}_R^{\pm\pm}} < M_{{\delta}_R^{\pm\pm}} + M_{\widetilde\chi_1^{0}}$. In this case the Higgsino will decay into the LSP and two same sign leptons through an off-shell doubly-charged Higgs boson. In Fig. \ref{fig:minv} we show the invariant mass distribution for the charged lepton pairs, where the doubly-charged Higgsino mass is 350 GeV, the
doubly-charged Higgs boson mass is 300 GeV and the LSP mass is 80 GeV. We see that in such a case the resonant peak in the same-sign charged lepton pair is lost but a kinematic edge exists at around 270 GeV. Note that
we still expect a narrow resonance from the direct pair production of the doubly-charged scalar and an
enhanced signal rate but we do not see any new enhancement at the resonance.

Experiments at the LHC are looking for doubly-charged Higgs bosons by analyzing final states with four high
$p_T$ charged leptons. Our model gives a resonant multi-lepton signal with large missing energy depending on the mass difference between the doubly-charged Higgs boson and the Higgsino. Such a signal accompanied by a peak in the same-sign lepton invariant mass distribution of the same-sign charged lepton pair. This will clearly
suggest an alternative signal not restricted to the direct production of doubly charged scalars.
This can definitely be a possible channel for the discovery of the doubly-charged Higgsinos which might be
worth looking for.

\vskip0.2in

\section{Discussion and Conclusion}
\label{sec:conclusion}

In this work we have studied the pair-production and decay of the doubly-charged Higgsinos in the left-right supersymmetric model and looked at the possible collider signatures at the LHC. The four lepton plus missing energy signal has a variety of distinct
features which can easily distinguish itself from other signals, arising especially from the minimal supersymmetric standard model.

We have studied the multi-lepton final state  $2\ell^+ 2\ell^- + \slashed{E}_T$ arising in the left-right SUSY model. We find that
there are three distinct sub-processes that contribute to the signal. We have shown through two
representative points in the model how each sub-process dominates the signal depending on the kinematic requirements on the missing transverse momenta. We also show through various kinematic distributions
the highlight of the four lepton signal in this model. Using specific cuts on the final states we find that there is
very little background from SM. The major background at the LHC where two $Z$ bosons decay into four
charged leptons is minimized by putting an invariant mass cut which neglects events at the $Z$ boson peak.
Thus, the signal produced by our model at the colliders would be clean and very easy to distinguish from
other competing models. Large missing transverse momenta in the final state can be triggered upon
to rule out contributions coming from the direct production of doubly-charged scalars and therefore
would give a strong hint of a supersymmetric model with doubly-charged particles.  The data collected by the LHC
experiments should already provide significant constraints on the masses of the doubly charged Higgsino and
Higgs boson through the process outlined here.  Dedicated searches for these doubly charged particles in the
channel proposed here by the experimental collaborations will be highly desirable.

\acknowledgments

KSB and AP are supported by the US Department of Energy Grant No. DE-FG02-04ER41036.
The work of SKR is partially supported by funding available from the Department of Atomic Energy,
Government of India, for the Regional Centre for Accelerator-based Particle Physics, Harish-Chandra
Research Institute.

\begin{appendix}
\begin{center}
{\bf APPENDIX}
\end{center}
In this Appendix we list down all the Feynman rules necessary for analyzing
productions and decays of doubly-charged Higgsinos in the LRSUSY model.


\noindent
{\bf\underline{Fermion-Fermion-Z Boson, $\gamma$:}}

$\displaystyle
\bullet \gamma^{\mu} \tilde\delta_{L,R}^{--} \bar{\tilde\delta}_{L,R}^{--}:~~~~~~~~~~~2i e \gamma^{\mu}$

$\displaystyle
\bullet Z_L^{\mu} \tilde\delta_{L}^{--} \bar{\tilde\delta}_{L}^{--}:~~~~~~~~~~~i\frac{g_L \cos 2 \theta_W}{\cos \theta_W} \gamma^{\mu}$

$\displaystyle
\bullet Z_L^{\mu} \tilde\delta_{R}^{--} \bar{\tilde\delta}_{R}^{--}:~~~~~~~~-i\frac{2g_L \sin^2 \theta_W}{\cos \theta_W} \gamma^{\mu}$

$\displaystyle
\bullet Z_R^{\mu} \tilde\delta_{L}^{--} \bar{\tilde\delta}_{L}^{--}:~~~~~~~~-i\frac{g_L \sin^2 \theta_W}{\sqrt{\cos 2 \theta_W}\cos \theta_W} \gamma^{\mu}$

$\displaystyle
\bullet Z_R^{\mu} \tilde\delta_{R}^{--} \bar{\tilde\delta}_{R}^{--}:~~~~~~~~~~~i\frac{g_L (1-3 \sin^2\theta_W)}{\cos \theta_W \sqrt{\cos 2 \theta_W}} \gamma^{\mu}$

$\displaystyle
\bullet Z_R^{\mu} u \bar{u}:~~~~~~~~~~~i\frac{g_L (3-8 \sin^2\theta_W+3\gamma_5 \cos 2\theta_W)}{12 \cos \theta_W \sqrt{\cos 2 \theta_W}} \gamma^{\mu}$

$\displaystyle
\bullet Z_R^{\mu} d \bar{d}:~~~~~~~~~~-i\frac{g_L (3-4 \sin^2\theta_W+3\gamma_5 \cos 2\theta_W)}{12 \cos \theta_W \sqrt{\cos 2 \theta_W}} \gamma^{\mu}$

\end{appendix}


\end{document}